\documentclass[preprint]{aastex}

\begin{document}

\title{Regimes Of Helium Burning}

\author{F. X. Timmes and J. C. Niemeyer}

\affil
{Center on Astrophysical Thermonuclear Flashes \\
 University of Chicago\\
 Chicago, IL 60637
}

\begin{abstract}

The burning regimes encountered by laminar deflagrations and ZND
detonations propagating through helium-rich compositions in the
presence of buoyancy-driven turbulence are analyzed. Particular
attention is given to models of X-ray bursts which start with a
thermonuclear runaway on the surface of a neutron star, and the thin
shell helium instability of intermediate-mass stars.  In the X-ray
burst case, turbulent deflagrations propagating in the lateral or
radial directions encounter a transition from the distributed regime
to the flamlet regime at a density of $\sim$ 10$^8$ g cm$^{-3}$. In
the radial direction, the purely laminar deflagration width is larger
than the pressure scale height for densities smaller than $\sim 10^6$
g cm$^{-3}$. Self-sustained laminar deflagrations travelling in the
radial direction cannot exist below this density.  Similarily, the
planar ZND detonation width becomes larger than the pressure scale
height at $\sim 10^7$ g cm$^{-3}$, suggesting that a steady-state,
self-sustained detonations cannot come into existance in the radial
direction.  In the thin helium shell case, turbulent deflagrations
travelling in the lateral or radial directions encounter the
distributed regime at densities below $\sim$ 10$^7$ g cm$^{-3}$, and
the flamelet regime at larger densities. In the radial direction, the
purely laminar deflagration width is larger than the pressure scale
height for densities smaller than $\sim 10^4$ g cm$^{-3}$, indicating
that steady-state laminar deflagrations cannot form below this
density. The planar ZND detonation width becomes larger than the
pressure scale height at $\sim 5\times10^4$ g cm$^{-3}$, suggesting
that steady-state, self-sustained detonations cannot come into
existance in the radial direction.

\end{abstract}


\section{Introduction}

Thermonuclear flames fronts are ubiquitous in stellar models of Type
Ia supernova, X-ray bursts, and shell flashes of intermediate stars.
Flame fronts propagating through carbon and carbon-oxygen compositions
are prevalent is in Type Ia supernova models (Timmes \& Woosley 1992;
Khokhlov 1995; Boisseau et al. 1996; Niemeyer \& Kerstein 1997), while
flame fronts propagating through helium compositions play a pivotal
role in X-ray burst models (Fryxell \& Woosley 1982; Taam 1987; Taam,
Woosley, \& Lamb 1996; Zingale et al. 2000) and shell flash models of
intermediate-mass stars (Schwarzschild \& H\"arm 1965; Paczynski 1974;
Iben 1977; Clayton \& De Marco 1997). In this paper we focus on flame
fronts propagating through helium compositions.

In most of the X-ray burst and thin shell flash models, the flame
front operates in a layer that is thin compared to the radius of the
star.  Local temperature perturbations in these thin layers become
amplified due to the nonlinear temperature dependence of the nuclear
reaction rates.  Ignition of the flame front probably takes place at
an individual point, or a small set of points, rather than
simultaneously throughout the entire layer as spherically symmetric
models require.  The flame front which subsequently propagates through
the thin layer may be a detonation, which travels faster than the
local sound speed, or a deflagration, which travels slower than the
local sound speed. Which type of flame front initially propagates
depends primarily on the prevailing thermodynamic conditions and
velocity fields.  Roughly speaking, the formation of a detonation is
suppressed in the presence of strong temperature fluctuations on
scales smaller than $\sim 10^3$ times the detonation width (Montgomery
et al. 1998).  The detonation or deflagration propagates both
laterally around the star and vertically through the thin layer. The
duration of a stellar event, or the observed rise time of the
luminosity, may be related in a fundamental way to the speed at which
the flame front propagates through the thin layer (e.g., Fryxell \&
Woosley 1982).

Neutron stars which accrete hydrogen-rich material from a binary
companion, or from the interstellar medium, end up with a thin layer
of helium-rich fuel on their surface.  The helium-rich fuel may be
spread evenly over the entire neutron star, or the fuel may be
confined to the polar regions if the magnetic field is sufficiently
strong.  For a relatively slow ($\dot {{\rm M}} \sim$ 10$^{-11}$
M$_{\sun}$ y$^{-1}$) accretion rate that is spherically symmetric, a
thick layer of helium ($\sim$ 10$^4$ cm) develops before ignition.
The density at the base of the accreted layer reaches $\sim$ 10$^8$ g
cm$^{-3}$ and the temperature reaches $\sim$ 10$^{8}$ K (Woosley \&
Taam 1976; Brown \& Bildsten 1998).  At these thermodynamic conditions
a planar detonation propagates in the lateral direction (perpendicular
to the direction of the gravitational force) at a 
speed of $S_{\rm D} \sim$ 10$^9$ cm s$^{-1}$, while a purely laminar
deflagration propagates in the lateral direction at a speed of $S_{\rm
L} \sim$ 10$^6$ cm s$^{-1}$ (Timmes 1999).  For larger accretion rates
(\hbox{$\dot {{\rm M}} \sim$ 10$^{-9}$ M$_{\sun}$ y$^{-1}$}), a thinner layer
of helium ($\sim$ 10$^3$ cm) is deposited before ignition.  The
density at the base of this accreted layer is $\sim$ 10$^6$ g
cm$^{-3}$ and the temperature reaches $\sim$ 10$^{8}$ K (Wallace,
Woosley, \& Weaver 1982; Brown \& Bildsten 1998).  For these
thermodynamic conditions, the lateral speed of a self-sustained
detonation is $S_{\rm D} \sim$ 10$^9$ cm s$^{-1}$, while a purely
laminar deflagration propagates in the lateral direction at a speed of
$S_{\rm L} \sim$ 10$^4$ cm s$^{-1}$ (Timmes 1999).  The vast
difference between these estimates for the planar detonation speeds
and the purely laminar deflagration speeds imply very different
predictions for the rise time of an X-ray burst (e.g., Fryxell \&
Woosley 1982).

The thin shell helium flash which occurs during the advanced
evolutionary stages of intermediate-mass stars (Schwarzschild \&
H\"arm 1965) results from nuclear burning being unable to raise the
overlying hydrogen envelope sufficiently to extinguish the reactions.
A drastic temperature rise at near constant pressure ensues. Shortly
after reaching its peak temperature, however, the material does expand
and cool on a rapid timescale.  Note that in contrast to the neutron
star example, the layer of fuel in the thin shell helium flash is
located inside a star rather than on the surface of a star.  We will
focus on a single model of a helium shell flash, namely, the M $\sim
1.0 {\rm M}_{\odot}$ core of a 7 M$_{\odot}$ star studied by Iben
(1977). Other models tested give qualitatively similar results.
Typical shell radii, densities and temperatures at the onset of this
thin shell helium flash model are \hbox{R $\sim 7\times10^8$ cm},
\hbox{$\rho$ $\sim$ 10$^4$ g cm$^{-3}$}, and T $\sim$ 2$\times$10$^8$
K, respectively. For these thermodynamic conditions, a self-sustained
detonation propagates in the lateral direction at $S_{\rm D} \sim$
10$^9$ cm s$^{-1}$, and a laminar deflagration propagates laterally at
$S_{\rm L} \sim$ 10 cm s$^{-1}$ (Timmes 1999).

Purely laminar deflagrations and planar ZND detonations (Zeldovich
(see Ostriker 1992); von Neumann 1942; D\"oring 1943) represent the
simplest one-dimensional, steady-state realizations of propagating
burning fronts.  Hydrodynamic instabilities in deflagrations, some of
which are intrinsic to three-dimensional propagating fronts and some
of which are related to the buoyancy of the hot burning products with
respect to the cold background in the radial direction, give rise to
growing perturbations and -- unless stabilized by non-linear effects
-- turbulence.  These instabilities deform the combustion surface,
potentially altering the structure of the reacting layers.  For
deflagrations, these instabilities generally increase the rate at
which nuclear energy is released. The deformed flame front propagates
at a speed faster than a purely laminar flame front as long as the
surface area increase dominates over the reduction of local energy
generation induced by hydrodynamic strain.  For a detonation, these
hydrodynamic instabilities generally do not increase either the energy
released from burning or the self-sustained detonation speed (e.g.,
Fickett \& Davis 1979).  Models for the non-linear evolution of
unstable or turbulent combustion fronts often involve statistical
tools, and the choice of an appropriate tool requires knowledge of the
burning regime of the combustion front.

The main purpose of this paper is to evaluate the regimes of helium
burning fronts in the presence of buoyancy-driven (convective)
turbulence.  A simple dimensional comparison of the relevant length
and speed scales of buoyancy-driven turbulence, ZND helium
detonations, and laminar helium deflagrations determine whether the
combustion front is in the flamelet regime, distributed burning
regime, or neither.  It will always be assumed in this paper that
convective stirring dominates over all other instabilities intrinsic
to the flame front. While cellular instabilities behind the detonation
front may be relevant in some contexts (Fickett \& Davis 1979;
Merzhanov \& Rumanov 1999), they will not be discussed in this paper.

\section{Relevant Scales}

The physical properties of laminar helium deflagrations, which are
primarily determined by a balance between nuclear energy generation
and the transport of internal energy, have been evaluated by Timmes
(1999) for a large grid of upstream densities and temperatures. We
will use the results of this survey for values of the laminar
deflagration speeds $S_{\rm L}$, widths $\delta$, and density
contrasts $\Delta \rho/\rho$ between the unburned fuel and its ash.
The density of the ash is smaller than the density of the unburned
fuel because the density declines behind a subsonic flame front.  

There are several plausible definitions for the width of the
deflagration. We consider three pragmatic definitions, each of which
can be measured by resolved calculations of laminar helium
deflagrations.  The first definition we consider is the distance
between where the temperature is 10\% above the upstream temperature
and where the nuclear energy generation attains its maximum
value. This width is called the reactive width and is denoted
$\delta_{{\rm nuclear}}$.  The second width definition we consider is
the distance between where the temperature is 10\% above the upstream
temperature and where the temperature reaches 90\% of its downstream
value.  This width is called the thermal width and is denoted
$\delta_{{\rm thermal}}$.  The third definition of a deflagration's
width that we consider is the distance between where the composition
has its upstream values and where the downstream composition first
reaches its final state. This width is called the composition width
and is denoted $\delta_{{\rm composition}}$. For laminar helium
deflagrations, the reactive widths $\delta_{{\rm nuclear}}$ are the
smallest widths. The thermal widths $\delta_{{\rm thermal}}$ are
usually slightly larger than the reactive widths $\delta_{{\rm
nuclear}}$, depending on the upstream thermodynamic conditions.  The
compositions widths $\delta_{{\rm composition}}$ are usually the
largest widths, ranging from being 1.2--10 times larger than the
thermal widths, depending mainly on the upstream density.  We will use
the $\delta = \delta_{{\rm nuclear}}$ in our analysis, but
qualitatively similar results are obtained if the differences in the
widths are taken into account.

The pressure scale height of a hydrostatically stratified helium layer
in some cases is comparable to the width of a helium combustion
front (e.g., Bildsten 1995), and should be included in any length
scale comparisons.  The pressure scale height is defined as
\begin{equation}
\label {pscale}
h_{\rm p} = {P \over \rho \ g}
\end{equation}
where $P$ is the scalar pressure, $\rho$ is the mass density, and $g =
G M/R^2$ is the acceleration due to gravity.  For the X-ray burst case
we assumed a $M = 1.4 M_{\odot}$ neutron star with a radius of
$R=10^6$ cm, while for the thin shell instability we the
aforementioned model from Iben (1977) that is characterized by $M \sim
1.0 M_{\odot}$, $R\sim7 \times 10^8$ cm.  The density in both X-ray
burst and thin shell instability cases is left as a free parameter,
permitting a classification of the helium burning regimes as a
function of density.

The amplitude of convective velocity fluctuations on large scales can
be estimated by evaluating the terminal rise velocity of buoyant
plumes with a diameter $\sim h_{\rm p}$ (Layzer 1955):
\begin{equation}
\label {vbuoy}
v_{\rm b} \approx 0.35 \ \left ( g \ {\Delta \rho \over \rho} \ h_{\rm p}
                         \right )^{1/2} 
= \ 0.35 \ \left( {\Delta \rho \over \rho \ \gamma} \right )^{1/2} \ v_{\rm s}
\enskip ,
\end{equation}
where $v_{\rm s}^2 = \gamma P/\rho$ is the local sound speed, 
and $\gamma$ is the ratio of specific heats.

For the purpose of classifying turbulent burning regimes, we are
interested mainly in the amplitude of turbulent velocity fluctuations
at the scale of the combustions front's width (Niemeyer \&
Kerstein 1997).  Assuming that the convectively driven, large scale,
fluctuations establish a turbulent cascade with the exponent $n$, we
can write
\begin{equation}
\label {vturb}
v_{\rm turb}(\delta) = \cases {
             v_{\rm b} \ \left ( {\delta / h_{\rm p}}\right ) ^n
             & {\rm if} $\delta < h_{\rm p}$ \cr
\noalign { \vskip .1 in }
             v_{\rm b}
             & {\rm otherwise.} \cr
                           }
\end{equation}

\noindent
The value of $n$ for buoyancy-driven turbulent cascades is not
unambiguously agreed upon. It appears reasonable at present to use
either $n = 1/3$ for Kolmogorov scaling, or $n = 3/5$ for
Bolgiano-Obukhov scaling (Niemeyer \& Kerstein 1997). We will assume
$n=1/3$ Kolmogorov scaling for the remainder of this paper.

\section{Helium burning regimes}

Deflagrations under astrophysical conditions are characterized by the
thermal diffusivity $\kappa$ dominating over all other microscopic
transport coefficients, such as viscosity $\nu$ and mass diffusivity
$D$.  In terms of the dimensionless numbers representing the transport
properties of the fluid, the Prandtl number $Pr = \nu/\kappa$ is very
small, and the Lewis number $Le = \kappa/D$ is very large.  As a
result of this disparity between the Prandtl and Lewis numbers the
conventional classification of turbulent burning regimes (e.g.,
Bradley 1993), which is based exclusively on time scale criteria, is
inappropriate (Niemeyer \& Kerstein 1997).  It is more appropriate in
such astrophysical conditions to combine the length scales and time
scales, and compare a turbulent diffusivity $D_{\rm turb}(l) \sim
v_{\rm turb}(l) \times l$ with the thermal diffusivity $\kappa$.  Note
the turbulent diffusivity is a function of the length scale being
examined. Since $D_{\rm turb}$ is a growing function of length scale
for most turbulent cascades, it is sufficient for our purposes to
consider the largest scale relevant for the flame structure, the flame
width $\delta$.  One may expect that a change of burning regimes
occurs when $D_{\rm turb} \sim \kappa$ or, equivalently, when $v_{\rm
turb}(\delta) \sim S_{\rm L}$.  It must be stressed that these
dimensional relationships may have potentially large dimensionless
coefficients that can only be determined by experiment or direct
numerical simulation.

The regime where $v_{\rm turb}(\delta) \ll S_{\rm L} \ll v_{\rm turb}(L)$ is
generally called the flamelet regime (Peters 1984; Clavin 1994).  This
regime is characterized by a nearly unperturbed laminar flame
structure on small scales and the deformation of the flame surface by
turbulent eddies on larger scales.  The growth of the flame surface
area as a result of turbulent wrinkling increases the energy deposited
by nuclear burning, which causes the deformed flame surface to
propagate faster than the purely laminar flame front.  This turbulent
flame speed $S_{\rm T}$ should scale linearly with the speed of the
fastest (and thus largest) turbulent eddies $v_{\rm turb}(L)$ if the
conditions for the flamelet region are to hold.  In the astrophysical
cases of interest here, a reasonable estimate is $L \approx h_{\rm
p}$.  A number of expressions for the turbulent flame speed $S_{\rm
T}$ as a function of the turbulent velocity fluctuations have been
proposed (e.g., Williams 1985; Yakhot 1988; Pocheau 1994; Shy et
al. 1996), but a well-founded consensus has yet to emerge.

The regime where $v_{\rm turb}(\delta) \gg S_{\rm L}$ is known as the
distributed burning regime (Peters 1984; Clavin 1994).  This regime is
characterized by a deformation of the flame surface even at small
scales.  Modeling attempts in this regime are severely hampered by the
overlapping length and time scales between the turbulent energy
transport and nuclear energy generation.  Damk\"ohler (1940) proposed
a simple re-normalization of the order-of-magnitude estimate for the
laminar flame.  This re-normalization assumes a vanishing influence of
the turbulence on the energy generation rate, which may be unrealistic
for distributed burning under astrophysical conditions.  Peters (1999)
gives an approach to extend flamelet regime models into the
distributed burning regime.

The length and speed scales of helium burning are shown in Figures
\ref{lengths} and \ref{speeds}, respectively. Each plot has the
upstream (unburned) mass density on the $x$-axis, the appropriate
scaling variable on the $y$-axis.  Each curve in the figure is for an
upstream composition of pure helium, and an upstream temperature of
10$^8$ K.  There are composition effects (the helium mass fraction)
and upstream thermodynamic effects (how cold the unburnt fuel is)
which determine the exact placement of each curve. However, the
magnitude of these two effects are small compared to the scale of the
$y$-axis and the order of the analysis.

The red curve in Figure \ref{lengths} shows the purely laminar flame
reactive width $\delta_{{\rm nuclear}}$ from the Timmes (1999)
survey. Curves for the thermal widths lie almost on top of the red
curve, while curves for the composition widths would be displaced
toward larger lengths for a given density.  The green curve gives the
pressure scale height for the X-ray burst case, while the blue curve
gives the pressure scale height for the thin shell case.  Both pressure
scale height curves are from equation (\ref{pscale}).

The red curve in Figure \ref{speeds} shows the laminar flame speed
from the Timmes (1999) survey. The green curve gives the buoyancy
speed from equation (\ref{vbuoy}).  Since the buoyancy speed
depends only on local thermodynamic conditions and the local laminar
flame properties, the buoyancy speed is identical for both the X-ray
burst and the thin shell instability cases.  The light blue curve
gives the turbulent speed at the scale of the laminar flame's reactive
width for the X-ray burst case, while the dark blue curve gives the
same quantity for the thin shell instability.  Both turbulent speed
curves, evaluated at the scale of the laminar flame reactive width,
are calculated from equation (\ref{vturb}).

Figure \ref{speeds} shows the curve for the thin shell turbulent speed
crosses the curve for the laminar flame speed at $\sim 10^7$ g
cm$^{-3}$.  At smaller densities the turbulent velocities on the scale
of the laminar flame width are larger than the speed of a purely
laminar deflagration, which corresponds to the distributed burning
regime.  In this case, turbulence must be expected to alter the
micro-structure of the deflagration flame front as the flame front
propagates in either the lateral or radial directions.  At densities
larger than $\sim 10^7$ g cm$^{-3}$, the laminar flame speed is larger
than the turbulent speeds on the scale of the laminar flame reactive
width, which corresponds to the flamelet regime. In this case, the
flame front which propagates in either the radial or lateral
directions has a nearly laminar deflagration structure on small scales
and a wrinkled surface on larger scales.  Turbulent deflagrations
travelling in the lateral or radial directions encounter the
distributed regime at densities below $\sim$ 10$^7$ g cm$^{-3}$, and
the flamelet regime at larger densities.

If the thermal deflagration widths are used instead of the reactive
widths, the cross-over density is nearly the same since the reactive
and thermal widths are of comparable magnitude.  If the deflagration
composition widths are used instead of the reactive widths, the
cross-over density is increased to $\sim$ 6$\times$10$^7$ g cm$^{-3}$,
since the composition width is generally larger then the reactive
width.  

Figure \ref{lengths} shows that the width of purely laminar
deflagration becomes larger than the pressure scale height at a
density of $\sim 10^4$ g cm$^{-3}$. This feature suggests that
steady-state laminar deflagrations travelling in the radial direction
cannot come into existance at densities smaller than $\sim 10^4$ g
cm$^{-3}$.  At the distributed regime to flamelet regime transition
density of $\sim 10^7$ g cm$^{-3}$, Figure \ref{lengths} indicates
that the laminar flame reactive width is much smaller than the
pressure scale height, suggesting validity of the steady-state
assumption for laminar flame propagation in either the lateral or
radial directions.

For the X-ray burst case, Figure \ref{speeds} shows that turbulent
speed curve crosses the laminar flame speed curve at $\sim 10^8$ g
cm$^{-3}$. Below this density a deflagration is in the distributed
regime, while at larger densities the deflagration is in the flamelet
regime.  If the thermal deflagration widths are used instead of the
reactive widths, the cross-over density is nearly the same.  If the
deflagration composition widths are used instead of the reactive
widths, the cross-over density is increased to $\sim$ 3$\times$10$^8$
g cm$^{-3}$, is not significantly larger than the reactive width at
these densities.

Figure \ref{lengths} shows that the width of purely laminar
deflagration becomes larger than the pressure scale height at a
density of $\rho \sim 10^6$ g cm$^{-3}$.  At densities smaller than
\hbox{$\rho \sim 10^6$ g cm$^{-3}$}, the deflagration structure in the
radial direction will look decisively different from steady-state
models, even without turbulence.  Steady-state laminar deflagrations
travelling in the radial direction cannot come into existance at
densities smaller than $\sim 10^4$ g cm$^{-3}$. While direct
simulations are needed to make specific predictions (see Bildsten 1995
for a modest attempt), one may expect that a deflagration propagating
in the lateral direction will broaden significantly in the radial
direction, and may never even reach a steady state.  At the
distributed regime to flamelet regime transition density of $\rho \sim
10^8$ g cm$^{-3}$, Figure \ref{lengths} indicates that the laminar
flame reactive width is much smaller than the pressure scale height,
suggesting validity of the steady-state assumption for laminar flame
propagation in either the lateral or radial directions.

For comparison with the turbulent deflagration case, we calculated the
structure of planar helium detonations under the ZND theory (Fickett \&
Davis 1979). Our results for the self-sustained detonation speeds and
thermodynamic conditions at the Chapman-Jouguet point are agree with
the values obtained by Mazurek (1973) and Khokhlov (1988, 1989).

There are several possible definitions for the width of a planar ZND
detonation. These include the distance from the shock front to the
point where the nuclear energy generation rate attains its maximum
value W$_{{\rm nucdot}}$, the distance from the shock front to the
point where the principle fuel has fallen to 1/10 of its initial value
W$_{{\rm composition}}$ (Khokhlov 1989), the distance from the shock
front to the point where 90\% of the total nuclear energy has been
releases W$_{{\rm nuclear}}$ (Khokhlov 1989), and the distance from
the shock front to where the composition reaches its final nuclear
statistical equilibrium state W$_{{\rm NSE}}$.  For planar helium
deflagrations, the widths defined by the energy generation rate
maximum W$_{{\rm nucdot}}$ are the smallest widths. Depending on the
upstream density, the composition widths W$_{{\rm composition}}$,
energy deposition widths W$_{{\rm nuclear}}$, and nuclear statistical
equilibrium widths W$_{{\rm NSE}}$ may be larger than the energy
generation rate maximum widths by factors of 1--15.  We will use the
W=W$_{{\rm nucdot}}$ in our analysis, but qualitatively similar
results are obtained if the differences in the various widths are
taken into account.

The detonation widths are shown by the purple curve in Figure
\ref{lengths}, and the detonation speeds are shown by the purple curve
in Figure \ref{speeds}. The speeds of self-sustained detonations, as
expected, are much larger than any other speed in Figures
\ref{speeds}. Perhaps suprisingly, the width of a self-sustained
detonations is larger than the width of a purely laminar deflagration
for any given density in Figure \ref{lengths}. For detonations, there
is a relatively long time between when material is first heated by the
passing shock wave, and when that material begins to burn
significantly (the induction time scale of detonations). For
deflagrations, there is a relatively short time between when material
is first heated by conduction and begins to burn significantly. These
time scales, when combined with the fact the speed of a self-sustained
detonation is supersonic while the speed of a purely laminar
deflagration is very subsonic, explain why the detonation widths are
larger than the deflagration widths in Figure \ref{lengths}.

For the X-ray burst case, Figure \ref{lengths} indicates that the
detonation width becomes larger that the the pressure scale height at
a density of $\rho \sim 10^7$ g cm$^{-3}$.  The steady state width is
larger than the radial ``box'' size containing the detonation.  This
suggests that a steady-state, self-sustained detonations cannot come
into existance in the radial direction at densities smaller than $\rho
\sim 10^7$ g cm$^{-3}$.  If the nuclear statistical equilibrium widths
are used instead of the reactive widths, the density at which the the
width becomes larger than the radial ``box'' size is increased to $\sim$
6$\times$10$^7$ g cm$^{-3}$.

For the thin shell case, the planar ZND detonation width becomes
larger than the pressure scale height at $\sim 5\times10^4$ g
cm$^{-3}$. Steady-state, self-sustained detonations travelling in the
radial direction cannot come into existance at densities smaller than
$\sim 5\times10^4$ g cm$^{-3}$.

\section{Summary}

\begin{figure}
\plotone{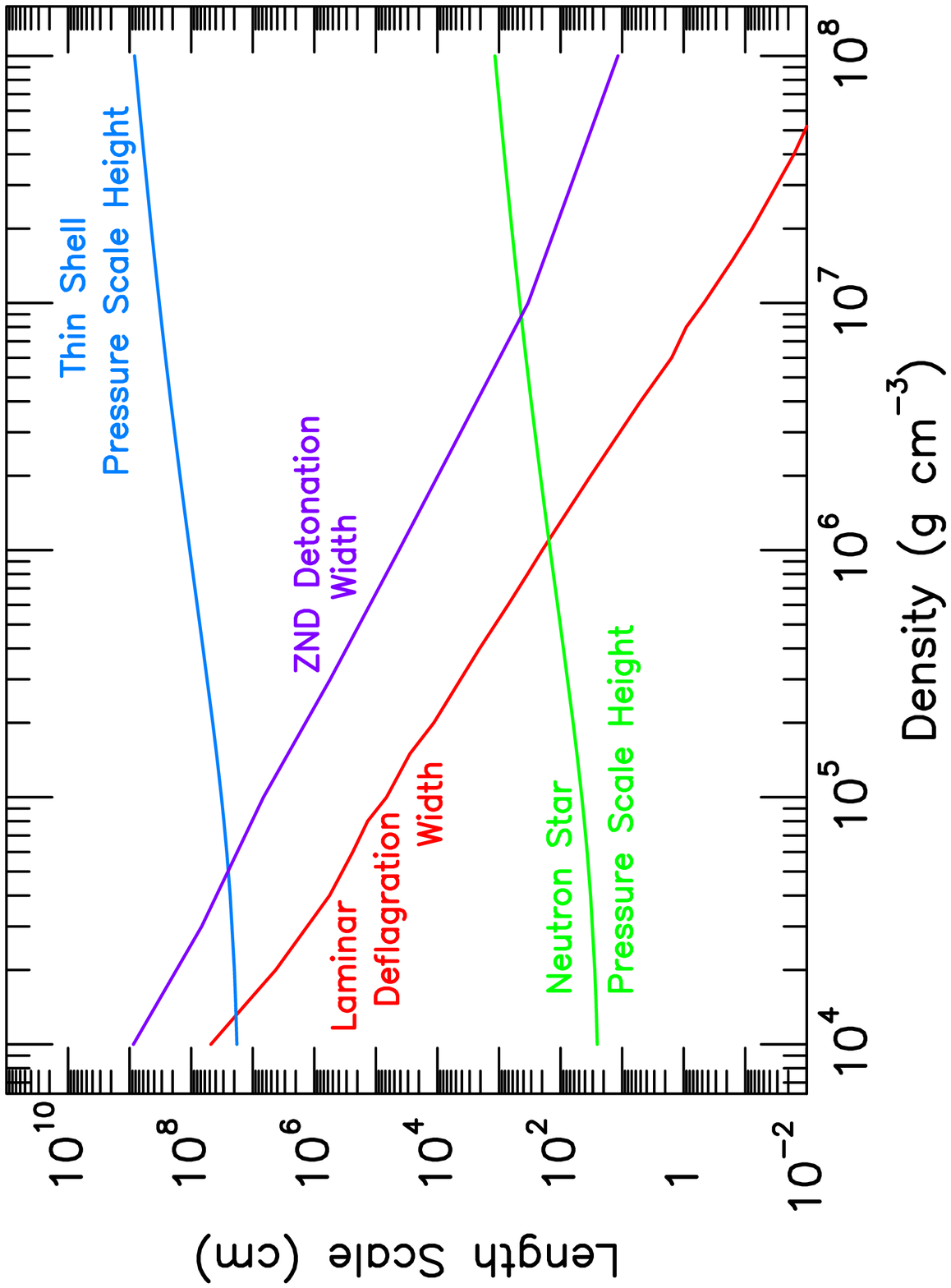}
\caption{\label{lengths} Length scales of helium burning.
The upstream (unburned) mass density on the $x$-axis, and 
the length scale variable is on the $y$-axis. }
\end{figure}

\begin{figure}
\plotone{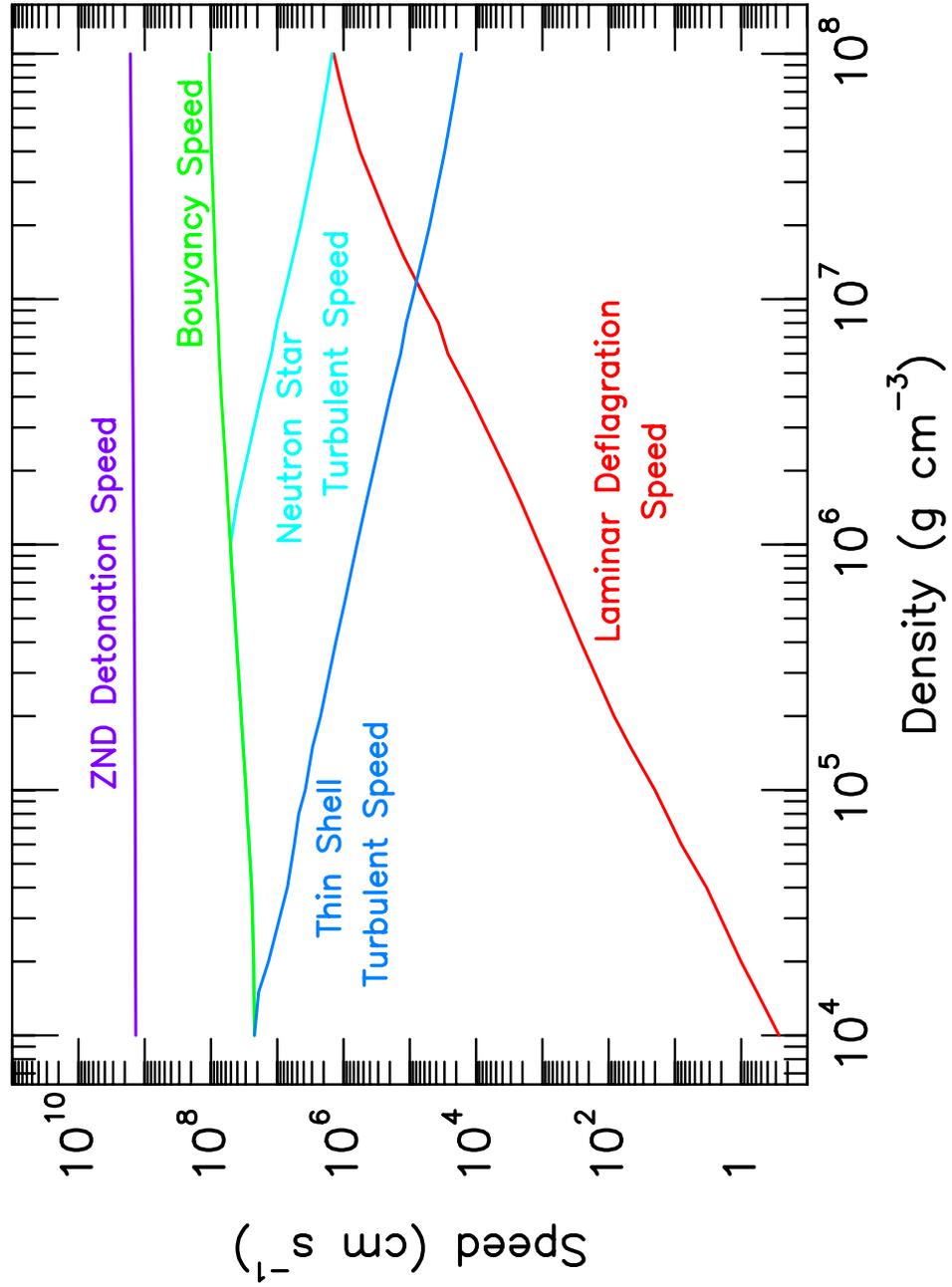}
\caption{\label{speeds} Speed scales of helium burning.
The upstream (unburned) mass density on the $x$-axis, and 
the speed scaling variable is on the $y$-axis.} 
\end{figure}

The various burning regimes encountered by deflagrations and
detonations propagating through helium-rich compositions in the
presence of buoyancy-driven turbulence have been analyzed, with the
main results being shown in Figures \ref{speeds} and \ref{lengths}.

For turbulent deflagrations in the X-ray burst case, there is a
transition from the distributed burning regime to the flamlet regime
at a density of $\sim$ 10$^8$ g cm$^{-3}$.  Turbulent deflagrations
propagating in either the lateral or radial directions will be
strongly deformed on micro-scales at smaller densities.  In addition,
the breakdown of the steady-state assumption for laminar deflagrations
propagating in the radial direction is signaled by the laminar
deflagration width becoming larger than the pressure scale height at
densities smaller than $\sim 10^6$ g cm$^{-3}$. This suggests that a
purely laminar deflagration cannot come into existence in the radial
direction at smaller densities. Similarly, the width of a
self-sustained, planar detonation is larger than the pressure scale
height at densities less than $\sim 10^7$ g cm$^{-3}$, indicating that
a steady-state detonation wave cannot come into existence in the
radial direction.

Turbulent deflagrations in the thin shell helium flash also encounter
both the flamlet and distributed regimes, with the cross-over density
being $\sim$ 10$^7$ g cm$^{-3}$.  Below this cross-over density, flame
fronts propagating in the radial and lateral directions are in the the
flamelet regime, where the flame front has nearly laminar deflagration
structure on small scales and a wrinkled surface on larger scales.  In
addition, the purely laminar deflagration width becoming larger than
the pressure scale height at densities smaller than $\sim 10^4$ g
cm$^{-3}$.  Purely laminar deflagration cannot come into existence in
the radial direction at smaller densities. Similarly, the width of a
planar detonation is larger than the pressure scale height at
densities less than $\sim 5\times10^4$ g cm$^{-3}$.  A steady-state
self-sustained detonation wave cannot come into existence in the
radial direction.

Figures \ref{lengths} and \ref{speeds} also have applications towards
helping to define what is meant by ``resolved'' numerical simulations
of helium burning.

\acknowledgements
This work has been supported by the Department of Energy under Grant
No. B341495 to the ASCI Center on Astrophysical Thermonuclear Flashes at
the University of Chicago.



\end{document}